\documentclass[11pt, a4paper]{article}
\usepackage{moriond,epsfig}

\def\be{\begin{equation}}
\def\ee{\end{equation}}
\def\bea{\begin{eqnarray}}
\def\eea{\end{eqnarray}}

\begin{document}
\vspace*{4cm}
\title{HADRONIC FINAL STATES AND SPECTROSCOPY\\IN EP COLLISIONS AT HERA}

\author{CARSTEN SCHMITZ\\for the H1 and ZEUS Collaborations}

\address{Universit\"at Z\"urich\\
Winterthurerstr.\ 190, CH-8057 Z\"urich\\
Switzerland}
\maketitle\abstracts{Recent results on spectroscopy and the measurement of hadronic final states
  in $ep$ collisions from the H1 and ZEUS collaborations are presented using data sets with an integrated luminosity between 44 and
  121 $\rm pb^{-1}$ collected during the HERA I running period. Besides a search for resonant states which could be interpreted as
  pentaquarks, a study of charged particle momentum spectra in the Breit frame
  and a measurement of neutral strange hadron production is shown. Furthermore
  two recent measurements of prompt photons are presented and compared with pQCD calculations. The
  measurements are performed in photoproduction ($\gamma p$) with
  a four-momentum transfer squared $Q^2\sim
  0\,\rm GeV^2$ or in deep inelastic scattering (DIS) at $Q^2>1 \rm \, GeV^{2}$.}

\noindent

{\small¥{\it Keywords}: pentaquark, prompt photon, charged particle
  multiplicity, spectroscopy, HERA.}

\section{Spectroscopy}
\subsection{Measurement of $K_S^0$, $\Lambda$ and $\bar{\Lambda}$ Production}
The production of the neutral strange hadrons $K_S^0$, $\Lambda$ and
$\bar{\Lambda}$ has been measured\,\cite{Chekanov:2006wz} 
by the ZEUS collaboration.
In addition to differential cross sections, further measurements are
presented, such as the baryon-antibaryon asymmetry, baryon-to-meson ratio, ratio
of strange-to-light hadrons, and the $\Lambda$ ($\bar{\Lambda}$) transverse
spin polarisation. 
\par
The ARIADNE
~Monte Carlo program
can describe the differential cross sections in transverse momentum $P_T$ and
pseudorapidity $\eta = - \ln \tan(\theta/2)$ of the hadrons
reasonably well when adjusting the strangeness-suppression factor to $\lambda_s=0.3$, although the cross section
at high $Q^2$ is overestimated.
ARIADNE adjusted to $\lambda_s=0.22$ is giving less
satisfactory results.
\par
The baryon-to-meson ratio $\mathcal{R}=(N(\Lambda)+N(\bar{\Lambda}))/N(K_S^0)$
varies between 0.2 and 0.5 in DIS, which is in agreement with measurements at $e^+e^-$ colliders\,\cite{Eidelman:2004wy}, where
$\mathcal{R}$ ranges between 0.2 and 0.4. 
$\mathcal{R}$ is described by ARIADNE
($\lambda_s=0.3$) to better than 10--20\%. In $\gamma p$ a dijet sample is
compared to the PYTHIA
~event generator. At large values of $x_{\gamma}^{\rm OBS}>0.75$,
where $x_{\gamma}^{\rm OBS}$ is a measure of the fraction of the photon energy transferred to the
dijet system, $\mathcal{R}$ is found to be roughly 0.4, which is in agreement with the
PYTHIA prediction and also corresponds to the values seen in DIS as discussed
above. For low values of $x_{\gamma}^{\rm OBS}$, 
$\mathcal{R}$ rises to a value of about 0.7, while PYTHIA predicts a flat $x_{\gamma}^{\rm OBS}$
dependence.
\par 
The measured ratio of strange-to-light hadrons
$\mathcal{T}=N(K_S^0)/(N(\pi^{\pm})+N(K^{\pm})+N(p)+N(\bar{p}))$ lies between
0.05 and 0.1 varying with $P_T$. A comparison with ARIADNE suggests a
strangeness-suppression factor $\lambda_s<0.3$.

\subsection{Search for Baryonic States Decaying to $\Xi\pi$ in Deep Inelastic Scattering}
A search for narrow baryonic resonances decaying into $\Xi^{-}\pi^{-}$ or
$\Xi^{-}\pi^{+}$ and their antiparticles is carried out\,\cite{H1_pentaquark} with the H1 detector
in DIS. 
No signal is observed for a new baryonic state in the mass range
1600--2300\,MeV 
in neither the doubly charged ($\Xi^{-}\pi^{-}$, $\bar{\Xi}^{+}\pi^{+}$) nor
the neutral ($\Xi^{-}\pi^{+}$, $\bar{\Xi}^{+}\pi^{-}$) decay channels.
In the neutral charged combinations there is a clear signal of the well-known
$\Xi(1530)^{0}$ resonance with 170 signal events.
The NA49 collaboration\,\cite{Alt:2003vb} observed two baryonic resonances
with masses around $1.86 \rm \, GeV$, which can be interpreted as pentaquark
states. The observation can not be confirmed by the H1 measurement.
\par
Mass dependent  upper limits at 95\% confidence level relative to the
$\Xi(1530)^{0}$ signal $R_{u.l.}(M)$ are derived
making use of a modified frequentist approach. 
The limits 
vary between 0.15
and 0.6 with values of $R_{u.l.}(1860) \sim 0.2(0.5)$ for the
doubly (neutral) charged combinations. The sum of all charged combinations
yields an upper limit of $R_{u.l.}(1860) \sim 0.5$, which is in agreement
with the upper limit of 0.29 derived by the ZEUS collaboration\,\cite{Chekanov:2005at}.

\subsection{Charged Particle Production in high $Q^2$ Deep Inelastic Scattering}
The process of parton fragmentation and hadronisation has been studied\,\cite{H1_scaledmomentum} by H1 using
charged particle momentum spectra at high $Q^2$. The measurement is performed
in the current region of the Breit frame with an energy scale given by $Q/2$.
Observables in the current region of the Breit frame can be compared with similar observables
measured in one hemisphere of $e^+e^-$ annihilation events, where the energy scale is
half the centre-of-mass energy $E^*/2$.
In the Breit frame the scaled momentum variable $x_p$ is
defined as $p_h^{\pm}/(Q/2)$, where $p_h^{\pm}$ is the momentum of a charged
track. In $e^{+}e^{-}$ annihilations the corresponding observable is $2p_h^{\pm}/E^{*}$.
\par
\begin{figure}[htb]
  \begin{center}
    \begin{minipage}[t][8cm][t]{0.9\textwidth}
      \begin{minipage}[t][8cm][b]{0.5\textwidth}
	\includegraphics[width=\textwidth]{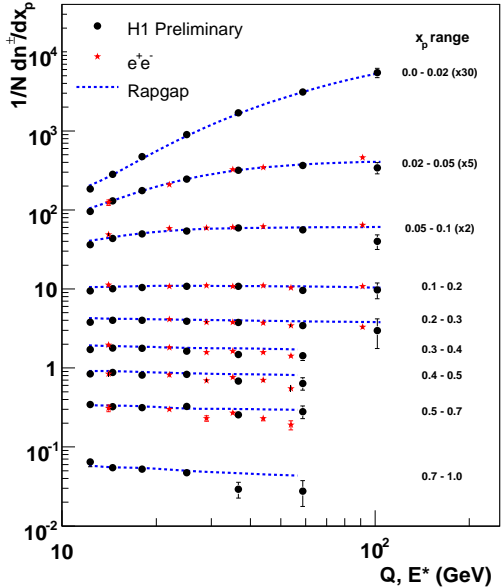}
      \end{minipage}
      \begin{minipage}[t][8cm][t]{0.5\textwidth}
	\vspace{0.5cm}
	\caption{H1 data for the event normalised inclusive scaled momentum
	  spectrum as function of Q for nine different $x_{p}$ regions compared
	  to data from $e^+e^-$ annihilation events. Also shown is the
	  prediction by the RAPGAP Monte Carlo program.}
	\label{fig:momentumspectrum}
      \end{minipage}
    \end{minipage}
  \end{center}
\end{figure}

The inclusive, event normalised charged
particle scaled momentum spectrum defined as 
\mbox{$D(x_p,Q)=(1/N_{event})dn^{\pm}/dx_p$} is shown in Fig.\
\ref{fig:momentumspectrum} as a function of $Q$ for
different bins of $x_p$ and is compared with results from $e^+e^-$
annihilation events. 
The $ep$ and $e^+e^-$ data are in
broad agreement supporting the concept of quark fragmentation
universality. The spectra become softer when moving from low to high Q, which
implies a substantial increase in the number of hadrons with a small share of
the initial parton's momentum. These scaling violations are assumed to be caused by parton
splitting in QCD, i.e.\ the same effect that causes the scaling violations
observed in deep inelastic structure functions.
The parton shower model as implemented in RAPGAP
gives the best description of the charged particle momentum spectra over the full range of $x_p$.

\subsection{Measurement of $\bar{d}$ and $\bar{p}$ Production in Deep Inelastic Scattering}
The production of (anti)deuterons, $d(\bar{d})$, and (anti)protons, $p(\bar{p})$, has been studied\,\cite{heavystable}
in DIS with the ZEUS detector and represents the first measurement of
$\bar{d}$ production in DIS. 
The (anti)deuterons and (anti)protons are identified by means of the energy-loss
measurement $dE/dx$ in the central tracking detector. 
\par
The corrected $\bar{d}$ production rate is found to be 3--4 orders of
magnitude lower than the corrected $\bar{p}$ yield, which is in agreement with
the H1 published data\,\cite{Aktas:2004pq} in $\gamma p$. Furthermore the
measured $\bar{p}/p$ ratio is consistent with unity. Within the given
statistics antitritons have not been observed.

\section{Prompt Photon Production}
\subsection{Inclusive Prompt Photon Production in Deep Inelastic Scattering}
A measurement of prompt photons in DIS\,\cite{H1_photonsDIS} has been presented by the H1
collaboration.
Compared to the previous measurement\,\cite{Chekanov:2004wr} of prompt photons in DIS, the
total cross section expectation is increased by roughly a factor of 10 due to
a markedly extended phase space.
The photon's transverse energy and pseudorapidity range is given by  $3<E_T^{\gamma}<10\,\rm GeV$ and $-1.2<\eta^{\gamma}<1.8$.   
\par
\begin{figure}[htb]
  \begin{center}
    \begin{picture}(160,180)
      \put(-150,0){\includegraphics[width=0.44\textwidth]{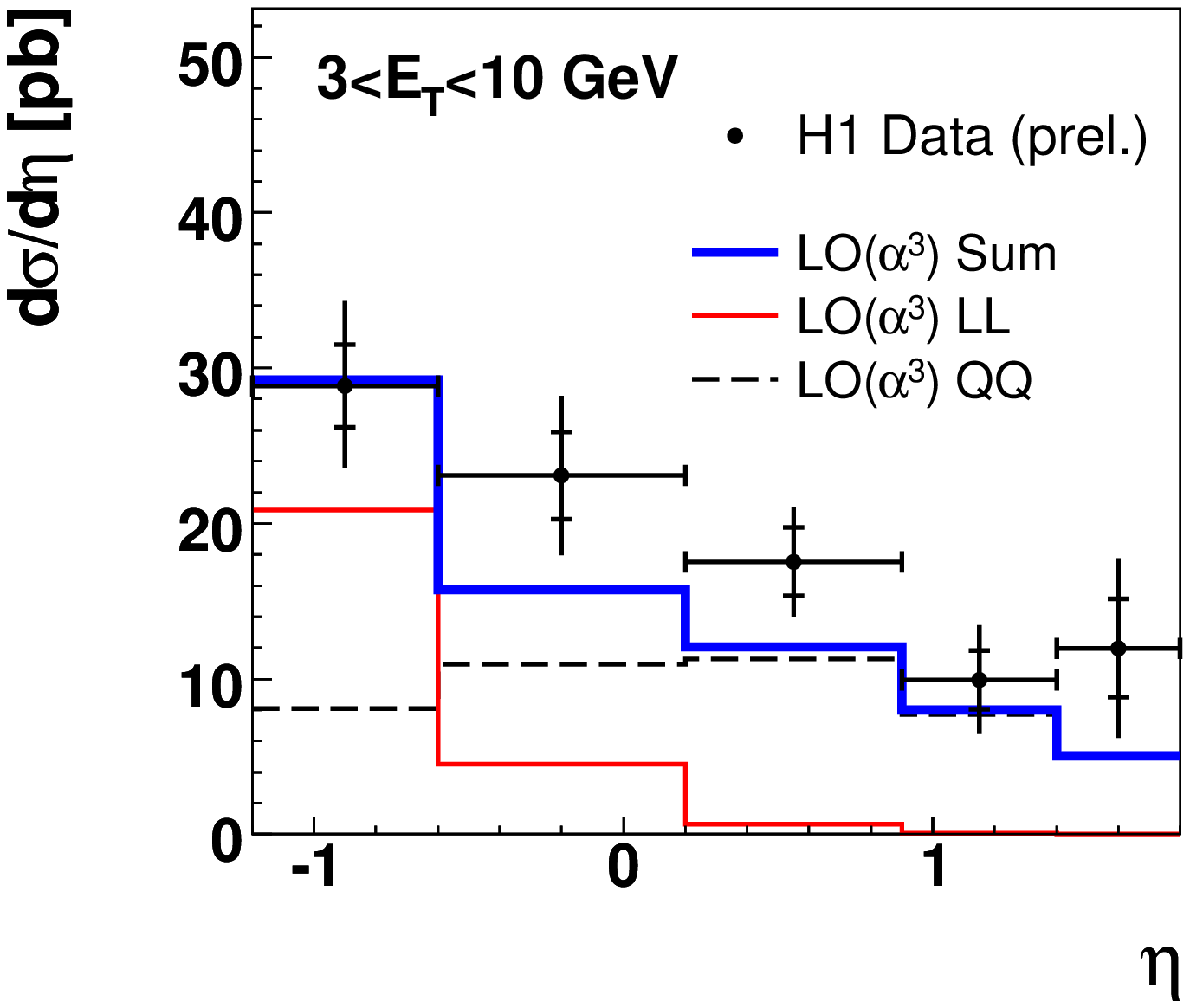}}
      \put(58,0){\includegraphics[width=0.55\textwidth]{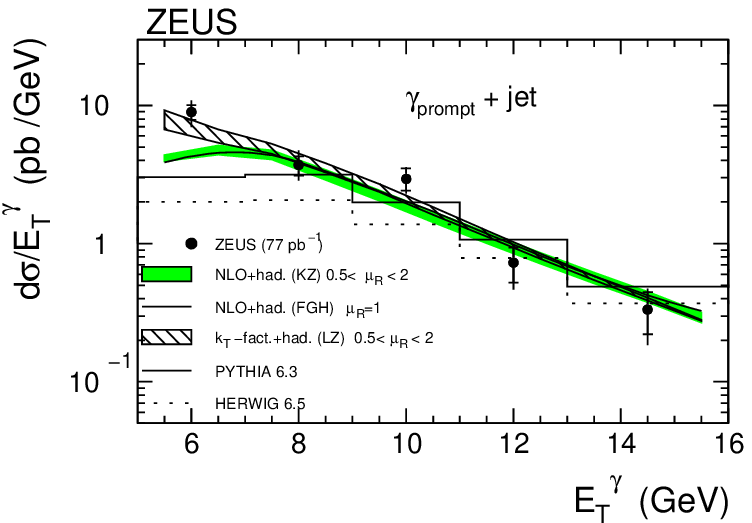}}
      \put(-104,165){\large \sf H1}
      \put(15,165){\sf (a)}
      \put(265,165){\sf (b)}
    \end{picture}
    \caption{Inclusive prompt photon differential cross section $d\sigma/d\eta^{\gamma}$
    in DIS as measured by H1 (a) and the photon plus accompanying jet
    differential cross section $d\sigma/dE_T^{\gamma}$ in $\gamma p$ as
    measured by the ZEUS collaboration (b). The measurements are compared to
    various pQCD calculations.} 
    \label{fig:photons}
  \end{center}
\end{figure}

Photons are identified by a compact electromagnetic cluster in the
main calorimeter with no track pointing to it. The major background due to
neutral hadrons inducing multi-photon clusters is considerably reduced by an
infrared-safe isolation criteria.
The extraction of the photon content from the remaining neutral hadron
background is done by a fit to the output of a multivariate shower shape analysis.
The measured differential cross sections $d\sigma/dE_T^{\gamma}$ and
$d\sigma/d\eta^{\gamma}$ are compared to a $\rm LO(\alpha^3)$
calculation\,\cite{Gehrmann-DeRidder:2006wz}. The differential cross section $d\sigma/dE_T^{\gamma}$ is well described in shape,
though the calculation is too low in normalisation. The underestimation is
most visible at central pseudorapidities as can be seen in Fig.\
\ref{fig:photons} a).
\par
At large pseudorapidities the dominant contribution is photon radiation by the
quark, while at low pseudorapidities, close to the scattered electron, the
contribution of photon radiation from the electron is dominant.

\subsection{Measurement of Prompt Photons with Associated Jets in Photoproduction}
The photoproduction of prompt photons, together with a separate jet in
addition to the photon, has been studied\,\cite{Chekanov:2006un} with the ZEUS
detector at HERA.
\par
The photon identification is based on the conversion probability of photons to
$e^+e^-$ in front of a preshower detector (BPRE). 
The photon kinematics is restricted to \mbox{$5<E_T^{\gamma}<16\,\rm
GeV$} and $-0.7<\eta^{\gamma}<1.1$, while the accompanying jet is selected in
the kinematic range \mbox{$6<E_T^{\rm jet}<17\,\rm GeV$} and \mbox{$-1.6<\eta^{\rm jet}<2.4$}.
A similar infrared-safe isolation criteria as in the above H1 analysis is used.
Differential cross sections as functions of $E_T^{\gamma}$ and $\eta^{\gamma}$
are compared to two NLO calculations (cf.\ references
in\,\cite{Chekanov:2006un}). As visible in Fig.\ \ref{fig:photons} b) both calculations describe
the data rather well, however underestimate the data at low $E_T^{\gamma}$. Another calculation (cf.\ reference in\,\cite{Chekanov:2006un}) based on a $k_T$ factorisation
approach yields the best description of the cross sections, particularly at
low $E_T^{\gamma}$.
When raising the minimum transverse energy to $E_T^{\gamma}>7\,\rm GeV$, the
data is well described by all three predictions.


\section*{References}

\begin{thebibliography}{99}
\bibitem{Chekanov:2006wz}
  S.~Chekanov {\it et al.}  [ZEUS Collaboration],
  arXiv:hep-ex/0612023.


\bibitem{Eidelman:2004wy}
  S.~Eidelman {\it et al.}  [Particle Data Group],
  Phys.\ Lett.\  B {\bf 592} (2004) 1.

\bibitem{H1_pentaquark}
  A.~Aktas {\it et al.}  [H1 Collaboration],
  H1-prelim-06-131, contributed paper to the 33rd International Conference on High Energy
  Physics, Moscow (2006). 
\bibitem{Alt:2003vb}
  C.~Alt {\it et al.}  [NA49 Collaboration],
  Phys.\ Rev.\ Lett.\  {\bf 92} (2004) 042003.
\bibitem{Chekanov:2005at}
  S.~Chekanov {\it et al.}  [ZEUS Collaboration],
  Phys.\ Lett.\  B {\bf 610} (2005) 212.
\bibitem{H1_scaledmomentum}
  A.~Aktas {\it et al.}  [H1 Collaboration],
  H1-prelim-06-033, contributed paper to the 33rd International Conference on High Energy
  Physics, Moscow (2006). 

\bibitem{heavystable}
  S.~Chekanov {\it et al.}  [ZEUS Collaboration],
  ZEUS-prel-06-008, contributed paper to the 33rd International Conference on High Energy
  Physics, Moscow (2006). 
\bibitem{Aktas:2004pq}
  A.~Aktas {\it et al.}  [H1 Collaboration],
  Eur.\ Phys.\ J.\  C {\bf 36} (2004) 413
\bibitem{H1_photonsDIS}
  A.~Aktas {\it et al.}  [H1 Collaboration],
  H1-prelim-06-031, contributed paper to the 14th International Workshop on
  Deep Inelastic Scattering, Tsukuba (2006).
\bibitem{Chekanov:2004wr}
  S.~Chekanov {\it et al.}  [ZEUS Collaboration],
  Phys.\ Lett.\  B {\bf 595} (2004) 86.
\bibitem{Gehrmann-DeRidder:2006wz}
  A.~Gehrmann-De Ridder, T.~Gehrmann and E.~Poulsen,
  Phys.\ Rev.\ Lett.\  {\bf 96} (2006) 132002;
  A.~Gehrmann-De Ridder, T.~Gehrmann and E.~Poulsen,
  Eur.\ Phys.\ J.\  C {\bf 47} (2006) 395.
\bibitem{Chekanov:2006un}
  S.~Chekanov {\it et al.}  [ZEUS Collaboration],
  Eur.\ Phys.\ J.\  C {\bf 49} (2007) 511.

\end{thebibliography}
\end{document}